\documentstyle[preprint,aps]{revtex}
\begin{document}
\preprint{\parbox[b]{3.3cm}{NIKHEF-94-P9\\NPL-1110\\hep-ph/9412287}}

\begin{title}
{Evolution of the spin of the nucleon}
\end{title}
\author{P.J.~Mulders
\thanks{Also at the Department of Physics and Astronomy, Free University,
Amsterdam}}
\address{ National Institute for Nuclear Physics and High Energy Physics
(NIKHEF).\\ P.O. Box 41882, NL-1009 DB Amsterdam, The Netherlands
\\Email address: pietm@nikhef.nl
}
\author{S.J.~Pollock}
\address{University of Colorado, Boulder CO, USA 80303
\\Email address: pollock@lucky.colorado.edu
}
\maketitle
\begin{abstract}
We compare momentum sum rules from unpolarized electroproduction and
the spin sum rule for $g_1$ in polarized electroproduction, and their
$Q^2$ evolution in the framework of the operator product expansion.
Second order effects in $\alpha_s$ are included.  We show that in
comparing the evolution of the spin sum rule with the momentum sum rule
one is not overly sensitive to using first or second order, even when
going to the extreme low $Q^2$ limit in which gluons carry no momentum.
Our results show that in that limit there is no need to include any
contribution of strange quarks.
\end{abstract}
\bigskip
\newpage

\section{Introduction}

Deep inelastic scattering (DIS) is an important tool for studying the
structure of hadrons. Through sum rules the experiments provide values
of specific quark and gluon operator matrix elements.
The framework for this is the operator product expansion (OPE)\cite{OPE}.
Experimentally measurable sum rules are expressed as the product of matrix
elements and coefficient functions.
Examples are the momentum sum rules measured in unpolarized deep inelastic
scattering and the Bjorken sum rule\cite{BJO} and Ellis-Jaffe sum
rule\cite{ELL} in polarized deep inelastic scattering \cite{EXP}.
Initial measurements of the latter, showing deviations from the Ellis-Jaffe
prediction, have been interpreted as an indication of a surprisingly
small contribution of the quark spin to the nucleon spin\cite{TH}.
One of the points relevant for the interpretation is the
scale dependence of the matrix elements and the coefficient functions,
which can be calculated in perturbation theory.
The QCD corrections to the Bjorken sum rule up to order
$\alpha_s^3$ have now been calculated in leading twist\cite{BETA,LARi,LARii},
and the higher twist corrections have
been estimated \cite{Twist}.  Recently, the order $\alpha_s^2$
corrections to the Ellis-Jaffe sum rule in leading twist and massless quark
approximation have been completed as well\cite{LARii}. These corrections
provide powerful means to further study the $Q^2$ evolution of the spin
structure of the nucleon.

For matrix elements that have no scale dependence deep inelastic measurements
immediately provide us with interpretable results that occasionally can
be compared with other experimental measurements in a completely different
domain, e.g. the Bjorken sum rule.
It is well known\cite{TH} that the singlet part of the first moment of
the spin structure function $g_1$
is not of this type and
has an anomalous $Q^2$ evolution. It
is also well known\cite{KOD} that the {\em leading} term in $\alpha_s$
in the axial anomalous dimension vanishes, and for this reason some
authors dismiss this $Q^2$ evolution as insignificant.  Roughly
speaking, corrections to the singlet first moment arising from the
anomalous dimension can be argued to behave like $\alpha_s \log Q^2$,
and hence appear {\em approximately} $Q^2$ independent.
In an earlier
paper,\cite{KUN} we argued that comparing momentum sum rules from
unpolarized electroproduction and the spin  sum rule for $g_1$,
including their $Q^2$ evolution, showed that DIS spin measurements are
{\em consistent} with a low energy valence quark picture, where the
valence quarks carry a substantial part of the spin of the proton, namely
of the order of $G_A$/(5/3) $\approx$ 0.75.
In this paper, we
extend our earlier calculations to fully include the next higher order
QCD corrections in leading twist.
In this way we are able to get a feeling for the sensitivity to the use
of first and second order in the evolution of the spin sum rule.
We also consider the possible effects of a strange quark threshold at very
low momentum scales.
With the results of more recent experiments we can give an estimate of the
``spin carried by quarks''.

\section{Formalism}

\subsection{The momentum sum rules}

As a typical example of an (unpolarized) ``quark momentum sum rule'',
the second moment of $F_2$ is given by
\begin{eqnarray}
\int_0^1\, dx F_2(x,Q^2)
&&= \sum_{i=1}^{n_f} e_i^2 \int\, x[q_i(x,Q^2)+\bar q_i(x,Q^2)]\  dx
\ \nonumber\\ &&
= \sum_{i=1}^{n_f} e_i^2 \epsilon_i(Q^2)
= \sum_{i=1}^{n_f} e_i^2\epsilon_i^{\rm NS}(Q^2) +
	\langle e^2\rangle \Sigma_{2}(Q^2),\label{eqii}
\end{eqnarray}
where $q_i(x,Q^2)$ is the quark distribution function and $\epsilon_i =
\epsilon_i(Q^2)$ is the momentum fraction carried by quarks and
antiquarks of flavor i ($n_f$ is the number of flavors), which is separated
into nonsinglet (NS) and singlet contributions. The quantity
$\Sigma_{2}\equiv\sum_i\epsilon_i$ is the total momentum
fraction carried by the (valence+sea) quarks, which can be expressed as
the matrix element of the quark part of the energy momentum tensor.
The quantity $\epsilon_i^{\rm NS}
\equiv \epsilon_i - \Sigma_{2}/n_f$ is the nonsinglet part of the
second moment for flavor i, and $\langle e^2\rangle $ is the average
quark charge.

There is no unique way to define parton distributions beyond leading
order, but we follow Buras' \cite{BUR} conventions, including
renormalization in the $\rm \overline{MS}$ scheme.
This results in the following formulae for the
(unpolarized) momentum sum rules, including next to leading order
corrections\cite{NLO}. For the (nonsinglet) valence quark momentum
sum rule
\begin{equation}
V_{2}\ =\ \sum_{i=1}^{n_f}\int_0^1\,x[q_i(x,Q^2) -\bar q_i(x,Q^2)]\ dx
\label{eqvb}
\end{equation}
the evolution is given by
\begin{equation}
V_2(Q^2) = \exp \left( - \int_{\alpha_s(Q_0^2)}^{\alpha_s(Q^2)} d\alpha^\prime
\, \frac{\gamma^{NS}(\alpha^\prime )}{2\beta (\alpha^\prime)} \right)
\, V_2 (Q_0^2),
\end{equation}
with the anomalous dimension given by
\begin{equation}
\gamma^{NS}(\alpha_s) = \gamma_0^{NS} \left( \frac{\alpha_s}{4\pi} \right)
+ \gamma_1^{NS} \left( \frac{\alpha_s}{4\pi} \right)^2 + \cdots,
\end{equation}
with $\gamma_0^{NS}$ = $64/9$ and $\gamma_1^{NS}$ = $96.6584-6.32\,n_f$.
The beta function governs the behavior of the strong coupling constant,
\begin{equation}
\mu^2 {d\alpha_s\over d\mu^2}\ = \  \beta(\alpha_s)\
        = \ -\beta_0 {\alpha_s^2\over 4\pi}
            -\beta_1 {\alpha_s^3\over 16\pi^2}-\cdots\ ,
\label{eqvi}
\end{equation}
with $\beta_0 = 11-2 \,n_f/3$, $\beta_1 = 102-38 \,n_f/3$.

The leading order (LO) solution for the strong coupling constant is
\begin{equation}
\frac{4\pi}{\beta_0 \alpha_s(Q^2)} - \log (Q^2) = \mbox{invariant},
\end{equation}
while the next to leading order (NLO) is
\begin{equation}
\frac{4\pi}{\beta_0 \alpha_s(Q^2)} - \log (Q^2)
- \frac{\beta_1}{\beta_0^2}\, \log \left( 1 + \frac{\beta_0^2}{\beta_1}\,
\frac{4\pi}{\beta_0 \alpha_s(Q^2)} \right) = \mbox{invariant}.
\end{equation}
We use these expressions to calculate the running coupling constant, i.e.
we make no further expansion.

The leading order solution for the valence quark momentum
sum rule $V_2(Q^2)$ using only the leading term in the $\gamma$-function reads
\begin{equation}
V_2(Q^2) = \left( \frac{\alpha_s(Q^2)}{\alpha_s(Q_0^2)}
\right)^{\frac{\gamma_0^{NS}}{2\beta_0}} V_2(Q_0^2).
\end{equation}
The next to leading order result reads
\begin{equation}
V_2(Q^2) = \left( \frac{\alpha_s(Q^2)}{\alpha_s(Q_0^2)}
\right)^{\frac{\gamma_0^{NS}}{2\beta_0}} \,
\left( \frac{4 \pi \beta_0 + \beta_1\alpha_s(Q^2)}
{4 \pi \beta_0 + \beta_1 \alpha_s(Q_0^2)}
\right)^{\frac{\beta_0\gamma_1^{NS}-\beta_1\gamma_0^{NS}}{2\beta_0\beta_1}}\,
V_2(Q_0^2).
\end{equation}
The NLO solution can be rewritten \cite{BUR} as the LO
solution times a polynomial in $\alpha_s$.
This is the result which we will refer to as truncated NLO. It reads
\begin{equation}
V_{2}(Q^2)\  = \ V_2(Q_0)^2
		\left[
		\alpha_s(Q^2)/\alpha_s(Q_0^2)
		\right]^{d_{\rm NS}^{\,(2)}}
		\left(1+{{\alpha_s(Q^2)-\alpha_s(Q_0^2)}\over 4\pi}Z_{\rm NS}
						\right),
			\label{eqiii}
\end{equation}
where the values for $d_{\rm NS}^{\,(2)}$ and $Z_{\rm NS}$ can be found in
Table \ref{tablei}.

For the singlet part the second moment of the distribution function for
quarks, $\Sigma_2(Q^2)$, mixes with the second moment of the gluon
distribution, $G_2(Q^2)$. One has, however, $\Sigma_2(Q^2)$ + $G_2(Q^2)$ = 1,
which makes it possible to write the evolution as
\begin{equation}
G_2(Q^2) = M^S(\alpha_s(Q^2)) \left[ G_2(Q_0^2) +
\int_{\alpha_s(Q_0^2)}^{\alpha_s(Q^2)} d\alpha^\prime
\frac{\gamma_{qq}(\alpha^\prime)}{2\beta (\alpha^\prime) M^S(\alpha^\prime)}
\right],
\end{equation}
with
\begin{equation}
M^S(\alpha_s) = \exp\left(-\int_{\alpha_s(Q_0^2)}^{\alpha_s} d\alpha^\prime
\, \frac{(\gamma_{qq} (\alpha^\prime )+\gamma_{GG} (\alpha^\prime )}
{2\beta (\alpha^\prime)} \right).
\end{equation}
Here $\gamma_{qq}(\alpha_s)$ and $\gamma_{GG}(\alpha_s)$ are
elements of the singlet anomalous dimension matrix,
expanded in $\alpha_s$ in the same way as the nonsinglet anomalous
dimension function.
For the second moment the anomalous
dimensions obey $\gamma_{qq} = -\gamma_{Gq}$ and $\gamma_{GG} = - \gamma_{qG}$.
The expansion coefficients are $\gamma_{qq,0}$ =
$64/9$, $\gamma_{qq,1}$ = $96.6584-10.2716\,n_f$, $\gamma_{GG,0}$ = $4 \,n_f/3$
and $\gamma_{GG,1}$ = $15.0864\,n_f$.

The solution for the function $M^S(\alpha_s)$, appearing
in the evolution of the singlet quark momentum sum rule
is in leading order given by
\begin{equation}
M^S(\alpha_s) = \left( \frac{\alpha_s}{\alpha_s(Q_0^2)}
\right)^{\frac{(\gamma_{qq,0}+\gamma_{GG,0})}{2\beta_0}},
\end{equation}
The NLO solution for $M^S(\alpha_s)$ is given by
\begin{equation}
M^S(\alpha_s) = \left( \frac{\alpha_s}{\alpha_s(Q_0^2)}
\right)^{\frac{(\gamma_{qq,0}+\gamma_{GG,0})}{2\beta_0}} \,
\left( \frac{4 \pi \beta_0 + \beta_1\alpha_s(Q^2)}
{4 \pi \beta_0 + \beta_1 \alpha_s(Q_0^2)}
\right)^{\frac{\beta_0(\gamma_{qq,1}+\gamma_{GG,1}) -
\beta_1(\gamma_{qq,0}+\gamma_{GG,0})}{2\beta_0\beta_1}}.
\end{equation}

This leads to the following LO solution for $G_2$,
\begin{equation}
G_2(Q^2) = \frac{\gamma_{qq,0}}{\gamma_{qq,0}+\gamma_{GG,0}}
+ \left( \frac{\alpha_s}{\alpha_s(Q_0^2)} \right)^
	{\frac{(\gamma_{qq,0}+\gamma_{GG,0})}{2\beta_0}} \,
\left[ G_2(Q_0^2) - \frac{\gamma_{qq,0}}{\gamma_{qq,0}+\gamma_{GG,0}} \right].
\end{equation}
For the NLO solution we do not have an analytic expression, but using the
result for $M^S$ a numerical solution is easily obtained. The truncated
NLO result is given by the coupled equations
\begin{eqnarray}
\Sigma_{2}(Q^2) &=& [\ (1-\tilde\alpha) \Sigma_{2}(Q_0^2) -
			 \tilde\alpha\, G_{2}(Q_0^2)]
		\left[
		\alpha_s(Q^2)\over\alpha_s(Q_0^2)
		\right]^{d_+^{\,(2)}}
		\left(1+{{\alpha_s(Q^2)-\alpha_s(Q_0^2)}\over 4\pi}Z_+\right)
			\nonumber \\
	 & &+ \tilde\alpha\,\left[1+\left\{{\alpha_s(Q_0^2)\over 4 \pi}
		\left[
		\alpha_s(Q^2)\over\alpha_s(Q_0^2)
	\right]^{d_+^{\,(2)}}-{\alpha(Q^2)\over 4\pi}\right\}K^\psi\right],
			\label{eqiv}
			\\
G_{2}(Q^2) &=& [-(1-\tilde\alpha)\ \Sigma_{2}(Q_0^2) +
			 \tilde\alpha\, G_{2}(Q_0^2)]
		\left[
		\alpha_s(Q^2)\over \alpha_s(Q_0^2)
		\right]^{d_+^{\,(2)}}
		\left(1+{{\alpha_s(Q^2)-\alpha_s(Q_0^2)}\over 4\pi}Z_+\right)
			\nonumber \\
	& & + (1-\tilde\alpha)\,\left[1+\left\{{\alpha_s(Q_0^2)\over 4 \pi}
		\left[
		\alpha_s(Q^2)\over\alpha_s(Q_0^2)
	\right]^{d_+^{\,(2)}} - {\alpha(Q^2)\over 4\pi}\right\}K^G\right].
			\label{eqv}
\end{eqnarray}
The $d$'s are the relevant anomalous dimensions for
this moment, here evaluated to first order. Higher order corrections
come from the $Z$'s and $K$'s, which are ($Q^2$ independent)
coefficients tabulated in table \ref{tablei}.  Note also that
conservation of momentum requires $\Sigma_{2}(Q^2)+G_{2}(Q^2) =
1$, which the above moments satisfy  due to the relation between
$\tilde\alpha, K^G$, and $K^\psi$.

\subsection{The singlet spin sum rule}

The singlet piece of the first moment of $g_1$ has recently been
computed to next to leading order in $\alpha_s$\cite{LARii}. This
includes both the singlet coefficient function $C^S$, as well as the
anomalous dimension, $\gamma^S$ of the singlet axial current in the
$\rm \overline{MS}$ scheme, using dimensional regularization. In the
adopted normalization, this yields for the sum rule
expressed in terms of the singlet axial matrix element
\begin{equation}
\Gamma_1^S(Q^2) =
\int_0^1\,g_1^S(x,Q^2) dx \ =\ C^S(\alpha_s(Q^2))\
\Delta\Sigma(Q^2),
\label{eqix}
\end{equation}
with
\begin{equation}
\Delta\Sigma(\mu^2) s_\sigma\ =\
   \langle p,s|J_\sigma^5|p,s\rangle \ =\
   \langle p,s|\sum_{i=1}^{n_f}\bar q_i \gamma_\sigma\gamma_5 q_i |p,s\rangle
	\ \equiv (\Delta u+\Delta d +\Delta s+\cdots) s_\sigma ,
\label{eqx}
\end{equation}
the quantity sometimes interpreted as the spin carried on the quarks.
The coefficient function
is given by
\begin{equation}
C^S(\alpha_s)\ =\ 1 - \alpha_s/\pi + \alpha_s^2/\pi^2\left(
				-4.5833+1.16248 \,n_f
						\right).
\label{eqviii}
\end{equation}
Not included in the sum rule for $g_1$ in Eq.~\ref{eqix} are higher twist
contributions.
The matrix element in Eq.~\ref{eqix} is scale dependent,
\begin{equation}
\Delta \Sigma (Q^2) = \exp \left( - \int_{\alpha_s(Q_0^2)}^{\alpha_s(Q^2)}
d\alpha^\prime \, \frac{\gamma^{S}(\alpha^\prime )}
			{2\beta (\alpha^\prime)} \right) \,
	\Delta \Sigma (Q_0^2),
\label{eq11}
\end{equation}
governed by the anomalous dimension, which with our normalization is
\begin{equation}
\gamma^{S}(\alpha_s) = \gamma_1^{S} \left( \frac{\alpha_s}{4\pi} \right)^2 +
\gamma_2^{S} \left( \frac{\alpha_s}{4\pi} \right)^3 + \cdots \label{eq12}
\end{equation}
with $\gamma_1^{S}$ = $16 \,n_f$ and
$\gamma_2^{S}$ = $314.67\,n_f-3.556\,n_f^2$.

The LO solution for the singlet axial charge $\Delta \Sigma$ reads
\begin{equation}
\Delta \Sigma (Q^2) = \exp \left(
\frac{\gamma_1^S}{8 \pi \beta_0} (\alpha_s(Q^2) - \alpha_s(Q_0^2))
				\right) \, \Delta \Sigma (Q_0^2),
\end{equation}
while the NLO solution reads
\begin{equation}
\Delta \Sigma (Q^2) =
\left(
\frac{4 \pi \beta_0 + \beta_1\alpha_s(Q^2)}{4 \pi
\beta_0 + \beta_1 \alpha_s(Q_0^2)} \right)^{
\frac{\beta_1\gamma_1^{S}-\beta_0\gamma_2^{S}}{2\beta_1^2}}\,
\exp \left(\frac{\gamma_2^S}{8\pi \beta_1} (\alpha_s(Q^2) -
\alpha_s(Q_0^2)) \right) \, \Delta \Sigma (Q_0^2).
\end{equation}
Finally, the truncated NLO solution for $\Delta \Sigma$ is
\begin{eqnarray}
\Delta \Sigma(Q^2) =
&&\biggl( 1 + \frac{\gamma_1^S}{8\pi \beta_0}
(\alpha_s(Q^2)-\alpha_s(Q_0^2))
+ \left(\frac{\beta_0 \gamma_2^S - \beta_1 \gamma_1^S}{64 \pi^2 \beta_0^2}
                \right) (\alpha_s^2(Q^2)-\alpha_s^2(Q_0^2))
\nonumber \\
&&\qquad\qquad\qquad\qquad\qquad
+ \frac{(\gamma_1^S)^2}{128 \pi^2 \beta_0^2}
 (\alpha_s(Q^2)-\alpha_s(Q_0^2))^2
\biggr) \Delta\Sigma(Q_0^2).
\end{eqnarray}

Experimental results are mostly given for $\Delta\Sigma (Q^2)$, which is
obtained from the experimental sum rule
by explicitly factoring out the coefficient function $C^S(\alpha_s(Q^2)$,
but {\em not} factoring out the $Q^2$ dependence in the matrix element,
given in Eq.~\ref{eq11}.
Note that $C^S(Q^2)$ differs at second order and beyond from the
analogous function $C^{NS}(Q^2)$,
\begin{equation}
C^{NS}(\alpha_s)\ =\ 1 - \alpha_s/\pi + \alpha_s^2/\pi^2\left(
			-4.5833 + n_f/3
						\right).
\end{equation}
which appears in the Bjorken sum
rule.  Being nonsinglet, the Bjorken sum rule has no analogous
anomalous dimension correction.

\section{Results}

In our previous work\cite{KUN}, we proposed fixing a quark model scale,
$Q_0^2$, where e.g. $G_{2}(Q_0^2)=0$, and/or
$V_{2}(Q_0^2)=1$.  This can be obtained by evolving from
experimental values at high $Q^2$\cite{MAR}. The spin sum rule was
considered in the same way, with a starting point $\Delta\Sigma(Q_0^2)$
taken from a quark model value, and then evolved up to $Q^2$ relevant
to DIS experiments.  In the bag model the estimate for
$\Delta\Sigma(Q_0^2) \approx 0.65$\cite{KUN}, the reduction from unity
coming from the lower components of the (relativistic) quark spinors,
the same source which reduces the axial charge in the bag model from
its nominal value of $5/3$.

The most important improvement of the results of \cite{KUN} is the inclusion
of the effects of corrections in the next order in $\alpha_s$.
This of course does not justify the use of perturbation theory in the
domain where we are using it, going to rather large values of the
strong coupling constant, $\alpha_s \sim 2$.
On the other hand, we can get some feeling for the convergence or
nonconvergence of our approach by comparing first and second
order evolution for the various moments.

It is well-known that for the running coupling constant the difference
between the first and second order results is large when one looks at
the functional dependence of $\alpha_s$ on $Q^2$. Similarly the evolution
of the moments as a function of $\alpha_s$
can also be strongly dependent
on the order.
Evolving down from $\alpha_s(M_Z^2) = 0.117$\cite{EXP,EK},
and the starting value $G_2(4$ GeV$^2) = 0.44$\cite{MAR},  using
{\em LO} order equations, gives $G_2 = 0$ when $\alpha_s = 1.80$.
Using {\em NLO} order equations gives $G_2 = 0$ when $\alpha_s = 1.79$.
Much larger is the difference for the valence momentum sum rule. Here
one finds that starting from $V_2(4$ GeV$^2) = 0.40$,  using
{\em LO} order equations, gives $V_2 = 1$ when $\alpha_s = 2.77$.
Using {\em NLO} order equations gives $V_2 = 1$ when $\alpha_s = 2.21$.

When we plot moments against each other as done in Fig~\ref{figi}, we
notice that the NLO calculations (dashed line) do not exhibit
a drastically different behavior as compared to the LO
calculations (solid). When we show NLO calculations,
the dashed line shows the exact solution to
the evolution equations. The dotted line shows the NLO
truncated expansions in $\alpha_s$ given in Eqs \ref{eqiii}, \ref{eqiv} and
\ref{eqv}.
The dot-dashed line is the same truncated expansion, but any terms involving
{\em higher} order corrections in $\alpha_s$ are evaluated using the
{\em leading} order
expression for $\alpha_s$, as suggested in ref~\cite{BUR}. Comparison of
the dotted, dot-dashed and solid line indicate typical uncertainties in
the NLO result. The differences between them is one order in $\alpha_s$
higher.

The same comparison of LO (solid) and three approximations for the NLO
results can be seen in Figs~\ref{figii} and~\ref{figiii}, which show
$\Delta\Sigma$ plotted against $V_2$ and $G_2$ respectively.
For the spin sum rule, we have compared our results with
$\Delta\Sigma(Q_{\rm exp}^2 = 10$ GeV$^2$) because we note from
Eq.~(\ref{eq11}) that the results remain proportional to the starting value
at all $Q^2$. This makes it easy to consider any scenario, e.g. starting
from a world average such as $\Delta\Sigma(4$ GeV$^2) = 0.31$\cite{EK} or
starting from a low-energy value\cite{KUN}.

In Fig. \ref{figi},
$\alpha_s$ is increasing down and to the right.  There is
no single value of $Q_0^2$ where both $V_2=1$ and $G_2=0$,
in either LO or NLO perturbation theory. In both cases
$G_2$ vanishes earlier (at higher $Q^2$), which is consistent
with an intuitive picture that at the quark model scale, meson-cloud
effects result in some residual $q \bar q$ sea.  Note that if one
plots e.g.  $V_{2}$ vs $\alpha_s$, there is a stronger
dependence on the order of perturbation theory used, as the evolution
of $\alpha_s$ is itself highly modified at these low $Q^2$.  It is
encouraging that when these observables are plotted against one
another, the trends are similar.
Furthermore, varying the value of $\alpha_S(M_Z^2)$ within current
experimental limits (e.g. using values ranging from 0.11 to 0.12)
has negligible effect on these curves.

In Fig. \ref{figii}, we show the relative value of the
singlet axial matrix element $\Delta\Sigma$ versus $V_{2}$, normalized to
the value at $Q_{\rm exp}^2$ = 10 GeV$^2$,
the characteristic scale of the SMC experiment.
The right edge corresponds to a value of $Q_0^2$ where $V_{2}(Q_0^2)=1$.
We see that $\Delta \Sigma(Q_0^2)/\Delta \Sigma(10)$
increases from about 1.71 (LO) to 1.98 (NLO).
This increase of the NLO evolution is, unlike that in Fig.~\ref{figi},
sensitive to the value of $\alpha_s(M_Z^2)$. Increasing the value
of $\alpha_s(M_Z^2)$ by 5\%, gives ratios of 1.83 (LO) and 2.21 (NLO).

In Fig. \ref{figiii},
$\Delta\Sigma$ is plotted against $G_{2}$, again normalized to its value
at $Q_{\rm exp}^2$ = 10 GeV$^2$.  Evolving from 10 GeV$^2$
to $Q_0^2$, this time fixed from $G_2(Q_0^2)=0$, gives for the ratio
$\Delta \Sigma(Q_0^2)/\Delta \Sigma(Q_{\rm exp}^2)$ values of
1.39 (LO) to 1.68 (NLO). The ratios are smaller than in Fig.~\ref{figii},
because $G_2 = 0 $ corresponds to a value of $V_2 < 1$.
Again a 5\% larger value of $\alpha_s(M_Z^2)$ results in
somewhat larger ratios, 1.45 (LO) and 1.85 (NLO).
The persistent enhancement of $\Delta \Sigma$ at the low-energy scale,
also in NLO, suggest that a valence picture with $\Delta \Sigma$ of the order
of 0.5 - 0.8 is consistent with the experimental result in DIS of the order
of 0.3 - 0.5. Clearly, the scale dependence cannot be neglected in
interpreting the results in deep inelastic experiments.

The evolution from $Q^2$ = 1 GeV$^2$ to 10 GeV$^2$ is presumably more reliably
in the perturbative regime, and in this case we find a ratio
$\Delta\Sigma(1)/\Delta\Sigma(10)$ of 1.031 (LO), or 1.068 (NLO) when
$\alpha_s(M_Z^2) = 0.117$. Especially here, we note a sensitivity
to $\alpha_s(M_Z^2)$. Taking its value to be 5\% higher, the ratios
becomes 1.040 (LO) and 1.110 (NLO).
The modification of the {\em actual} singlet moment $\Gamma_1^S$,
which includes $C^S(Q^2)$ as well, shows a decrease for $\Gamma_1^S$
when going to lower momenta, the ratios being
$\Gamma_1^S(1)/\Gamma_1^S(10)$ = 0.980 (LO) and 0.956 (NLO).
(Taking
$\alpha_s(M_Z^2)$ 5\% higher gives 0.973 (LO) and 0.921 (NLO).)
Although fairly small
in this region, the contribution to the evolution from the anomalous
dimension clearly can and should be taken into account, and goes beyond
the standard QCD effect arising purely from $C^S(Q^2)$, which is
sometimes all that is taken into account. Note furthermore that in
$\Gamma_1^S(Q^2)$, higher twist contributions proportional to $1/Q^2$
could contribute. These have not been included in the above estimate for
$\Gamma_1^S$, which refers purely to the twist two part.

Another point that deserves discussion is the inclusion of the effects
of the strangeness threshold. If one considers the $K\overline K$
threshold, i.e. $Q^2 \approx$ 1 GeV$^2$, as an appropriate value, a
large part of the evolution down to $Q_0^2$ involves $n_f$ = 2. If the
strangeness content of the nucleon has not become zero, the decoupling
of strangeness from the evolution leads to ambiguities in the
treatment. It would require a global analysis, which takes carefully
into account existing inequalities such as $\Delta s(x)$ $\le s(x)$ and
the consequences for the moments. The evolution equations with $n_f =
2$ instead of $n_f = 3$ in general tend to somewhat slow down the
increase of $\Delta\Sigma$ at lower momentum scales.

Finally, if we assume that (i) $\Delta\Sigma(Q_0^2) = 0.65$ as an appropriate
value at the low momentum scale, e.g. from an effective low energy
model\cite{KUN},
and (ii) pQCD continues to work down to low $Q^2$, and (iii) the asymptotic
values of the nonsinglet
combinations of the axial matrix elements are known from weak decays,
$\Delta u - \Delta d$ = 1.257 and (from low energy hyperon decays)
$\Delta u +\Delta d - 2 \Delta s = 0.58 \pm 0.05 $,
then we are able to calculate at any scale the axial matrix elements
for each of the quark flavors. The values at $Q_0^2$ (corresponding to
$G_2$ = 0), 1 and 10 GeV$^2$ are given in Table~\ref{tableii}. Assuming
three active flavors at $Q_0^2$ gives a positive value for $\Delta_s$.
In this case we have the possibility to incorporate the strangeness threshold
in a natural way, using only two active flavors to evolve $\Delta \Sigma$
down to 0.58, then continuing with three active flavors. The numbers in
this scenario for NLO are given in Table~\ref{tableii}. The strangeness
threshold required is $Q_s^2$ = $0.286$ GeV$^2$. We note that decoupling of
the strange quarks in the momentum sumrule at this same threshold implies
that at 4 GeV$^2$ the momentum carried by the strange antiquarks as compared
to nonstrange antiquarks is $2\,\overline{s}_2/(\overline{u}_2 +
\overline{d}_2)$ = 0.57.

In conclusion, using NLO
equations, a very satisfactory picture is obtained running all the way
from a low-energy-scale proton without gluons but with some nonstrange sea,
acquiring nonzero values for strangeness matrix elements only starting at the
strangeness threshold which is slightly above the scale where $G_2$ = 0.
We have analyzed the errors arising from an uncertainty in the strong
coupling constant
$\alpha_s(M_Z)$ = $0.117 \pm 0.005$ and those coming from an uncertainty in
the octet axial charge, $0.58 \pm 0.05$.
These are indicated in Table~\ref{tableii}. Note that
the results for $\Delta u$ and $\Delta d$ are not sensitive to the octet
axial charge if this is taken to coincide with the strangeness threshold.
Using these results and including second order QCD corrections everywhere,
we find that
\begin{eqnarray}
&\Gamma_1^p(10\ {\rm GeV}^2) &= (0.109 \pm 0.002) +
			(0.062 \pm 0.007 \pm 0.009) \Delta
\Sigma (Q_0^2) \nonumber \\
&&= 0.149 \pm 0.006 \pm 0.006, \\
&\Gamma_1^n(10\ {\rm GeV}^2) &= (-0.080 \pm 0.001) +
			(0.062 \pm 0.007 \pm 0.009 ) \Delta
\Sigma (Q_0^2) \nonumber \\
&&= -0.040 \pm 0.003 \pm 0.006, \\
&\Gamma_1^d(10\ {\rm GeV}^2) &\equiv 0.5 (\Gamma_1^p+\Gamma_1^n)
(1-1.5\,\omega_D) \nonumber\\
&&= (0.013 \pm 0.001) + (0.056 \pm 0.006 \pm 0.008)
 \Delta \Sigma (Q_0^2) \nonumber \\
&&= 0.049 \pm 0.004 \pm 0.005
\end{eqnarray}
(using the usual D-state admixture of 6\% in the latter).  The first
error in each term arises from our assumed uncertainties in
$\alpha_s(M_Z)$ and from the octet part of the sum rule, here added
in quadrature.  The second error bar (if shown) comes from our estimate
of the prescription dependence associated with evolving $\Delta\Sigma$
from 10 GeV$^2$ to $Q_0^2$.  This includes e.g. differences in
truncation schemes (see Figs. 2 and 3),  choice of s-quark threshold
mechanism, and determination of $Q_0^2$ from $V_2=1$ rather than $G_2=0$.
These have all been discussed above. We conservatively estimate $\Delta
\Sigma(Q_0^2)/\Delta \Sigma(10)$ = $ 1.65 \pm 0.25$, and the final
numbers above correspond to this choice, with $\Delta \Sigma(Q_0^2) =
0.65$.
The relations between the experimental sum rules and the "spin carried
by the quarks", i.e. $\Delta \Sigma (Q_0^2)$ are illustrated in
Fig.~\ref{figiv}.

\section{Conclusions}

In this paper we have investigated the extent to which measurements of
the spin sum rule at high energies should be interpreted, in view of
the role of their scale dependence. We have investigated the spin sum
rule together with the momentum sum rules in a systematic way,
extending our earlier results that used purely leading order evolution
to results that use next to leading order evolution. This has become
possible in part due to the recent work of Larin\cite{LARii}.  We have
estimated uncertainties arising from scheme dependence and higher order
QCD effects in the evolution, by using several truncation
prescriptions.  Our results indicate that many qualitative features
present in the leading order remain the same. Quantitative differences
show up, but do not upset the picture. In particular, we have
considered momentum sum rules for valence quarks and gluons compared
with one another or spin sum rules compared with the momentum sum
rules.  In general, we have not made any {\em interpretation} of the
relative gluonic contribution to the results for the singlet moment,
which helps avoid obvious scheme-dependent assumptions.  (see, e.g.
reference \cite{Ell}, which shows that the gluon contribution is zero
in certain renormalization schemes). We are simply using the operator
product expansion for the moments of interest, at next to leading
order.

The results for evolution from 1 GeV$^2$ to 10 GeV$^2$ are quite
reliable, and the fractional change evolving down to $Q_0^2$ is
apparently reasonably stable to next to leading order corrections too.
Of course, the moderate sensitivity in going from first to second order
is no proof of the reliability of the perturbative expansion.  We have
noted that while the sensitivity to using first or second order is
moderate, the sensitivity to the value of $\alpha_s(M_z)$ is quite
strong. We find that the inclusion of a strangeness threshold is not
very important for the rate of evolution.  Much more important is the
fact that there is a large change in $\Delta\Sigma$ running from 10
GeV$^2$ to a low-momentum scale $Q_0^2$, large enough that it can
easily reach a point where $\Delta \Sigma$ = $\Delta u + \Delta d -
2\Delta s$, i.e.  $\Delta s$ = 0, a natural point for the strangeness
threshold.  We have made this more explicit by starting with a
value\cite{KUN} of $\Delta \Sigma (Q_0^2)$ = 0.65, although any
assumption that one can estimate $\Delta\Sigma(Q_0^2)$ from a low
energy nucleon model is very clearly subject to debate. This scenario
implies that at present there is no reason to require an anomalously
large strange contribution in the proton at a low-momentum scale. Of
course, independent measurements of the strangeness content at this
scale (e.g.  from elastic neutrino - nucleon scattering) are important
to confirm this.

\section{ACKNOWLEDGMENTS}

This work is supported in part (S.J.P) by U. S.  Department of Energy grant
DOE-DE-FG0393DR-40774 and in part (P.J.M.) by the foundation for Fundamental
Research
of Matter (FOM) and the National Organization for Scientific Research
(NWO). SJP acknowledges the support of a Sloan Foundation Fellowship.

\newpage
\begin{figure}
\caption{Plot of $G_2$ vs $V_2$. Solid (dashed) curve shows
first(second) order calculations. We started from experimental
values\protect\cite{MAR} at roughly 4 GeV$^2$ and evolve downwards.
Dotted curve is 2nd order, but truncated. Dash-dotted is again 2nd order,
truncated, but with $\alpha_s$ in all higher order terms replaced with
its leading order expression.
}
\label{figi} \end{figure}

\begin{figure}
\caption{Same as Fig. 1, but plotting $\Delta\Sigma/\Delta\Sigma(Q_0^2)$
 versus $V_2$.
Curves are labeled as before.
}
\label{figii} \end{figure}

\begin{figure}
\caption{Same as Fig. 3, but plotting $\Delta\Sigma/\Delta\Sigma(Q_0^2)$
versus $G_2$.}
\label{figiii} \end{figure}

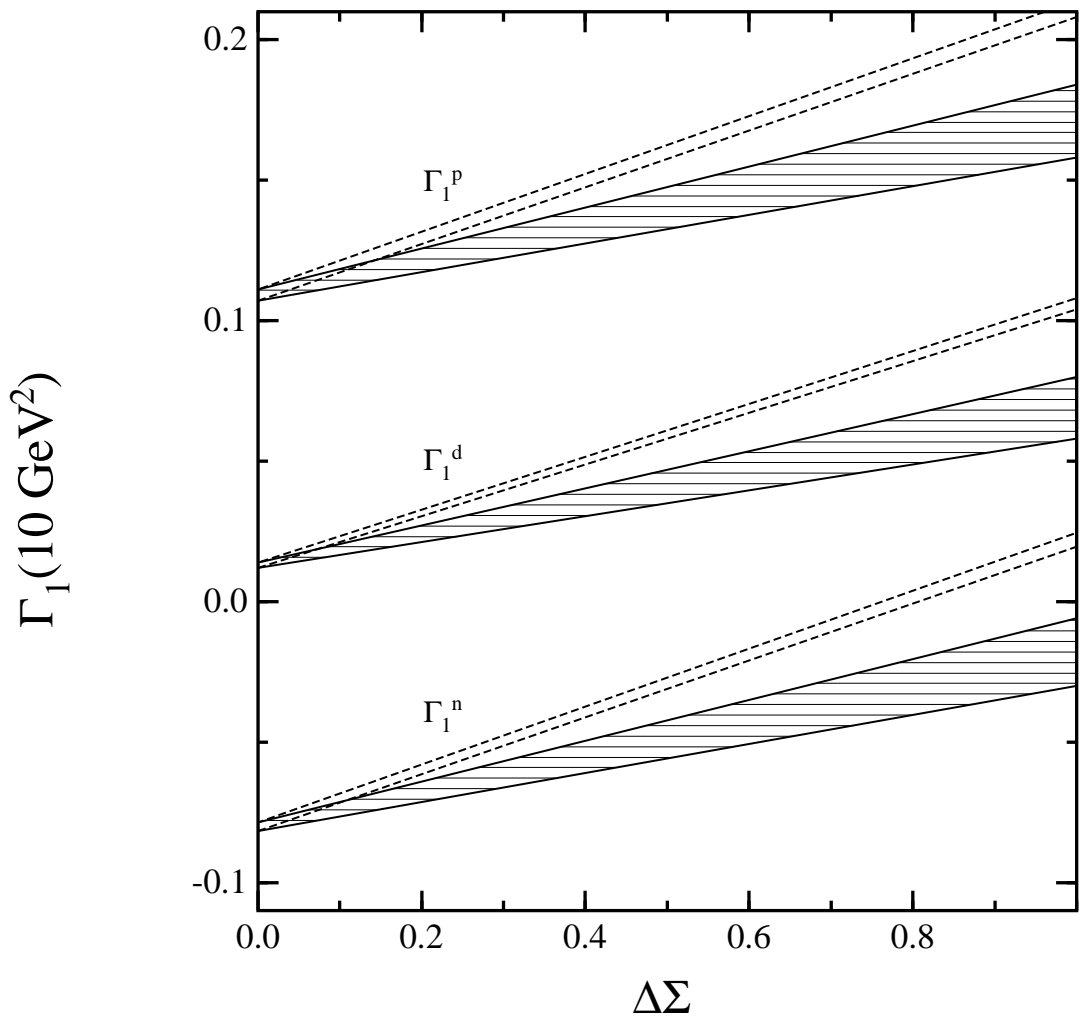
\begin{figure}
\caption{Experimental sum rules as a function of the spin carried by the
quarks, $\Delta\Sigma(Q_0^2)$, including uncertainties from
$\alpha_s$, octet axial charge, and scheme dependence are given by the
shaded areas.
The areas enclosed by the dashed lines are the results as a function of
$\Delta \Sigma (10\ {\rm GeV}^2)$.}
\label{figiv} \end{figure}

\widetext
\begin{table}
\caption{ Numerical values \protect\cite{BUR,NLO}
used for
the various parameters appearing
in the truncated next to leading order solutions for the momentum sum rules,
Eqs~\protect\ref{eqiii} through \protect\ref{eqv}.
}
\bigskip
\begin{tabular}{ccccccccccc}
$n_f$ & $\tilde\alpha$ & $d^{\,(2)}_{\rm NS}$ & $d^{\,(2)}_+$
	& $Z_+$ & $Z_{\rm NS}$ & $K^\psi$ & $K^G$ \\
\tableline
2  &  0.2727 & 0.3678 & 0.5058 & 1.486 & 1.428 & 0.4544 & -0.1704  \\
3  &  0.36   & 0.3951 & 0.6173 & 1.783 & 1.507 & 2.121 & -1.193  \\
4  &  0.4286 & 0.4267 & 0.7467 & 2.355 & 1.654 & 5.895 & -4.421  \\
5  &  0.4839 & 0.4638 & 0.8986 & 3.341 & 1.904 & 22.604& -21.191 \\
\end{tabular}
\label{tablei}
\end{table}

\begin{table}
\caption{ Values for the axial matrix elements of the quarks
using a starting value of $\Delta\Sigma(Q_0^2)$ = 0.65 at the
scale where $G_2(Q_0^2)$ = 0 using NLO results and a strangeness
threshold at the point where $\Delta s$ = 0. The errors refer to
uncertainties in $\alpha_s(M_Z)$ = 0.117 $\pm$ 0.005,
and in the octet axial charge, 0.58 $\pm$ 0.05 (underlined errors).
}
\bigskip
\begin{tabular}{ccccc}
$Q^2$ [GeV$^2$] & $\Delta u$ & $\Delta d$ & $\Delta s$ & $\Delta \Sigma$ \\
\tableline
$Q_0^2$ & 0.954   & -0.304   & - & 0.650 \\
$Q_s^2$ = 0.3 $\mp$ 0.1  & 0.919 $\pm$ \underline{0.025} & -0.339
$\pm$ \underline{0.025} & 0.00  & 0.580 $\pm$ \underline{0.05} \\
1 & 0.873 $\mp$ 0.01 & -0.384 $\mp$ 0.01 &
-0.045 $\mp$ 0.01 $\mp$ \underline{0.025} & 0.445 $\mp$ 0.03 $\mp$
\underline{0.025} \\
4 & 0.866 $\mp$ 0.01 & -0.391 $\mp$ 0.01 &
-0.052 $\mp$ 0.01 $\mp$ \underline{0.025} & 0.423 $\mp$ 0.03 $\mp$
\underline{0.025} \\
10 & 0.864 $\mp$ 0.01 & -0.393 $\mp$ 0.01  &
-0.055 $\mp$ 0.01 $\mp$ \underline{0.025}  & 0.417 $\mp$ 0.03 $\mp$
\underline{0.025}  \\
\end{tabular}
\label{tableii}
\end{table}


\end{document}